\documentclass{iopjournal}

\usepackage{amsmath}   
\usepackage{amssymb}
\usepackage{ulem} 
\usepackage{CJKutf8}

\begin{document}

\articletype{Paper} 

\title{Third Quantization for Order Parameter (I): BCS–BEC crossover with macroscopically coherent state}

\author{Guo-Jian Qiao (\begin{CJK}{UTF8}{gbsn}乔国健\end{CJK})$^1$, Miao-Miao Yi (\begin{CJK}{UTF8}{gbsn}易淼淼\end{CJK})$^1$,  Xin Yue (\begin{CJK}{UTF8}{gbsn}岳鑫\end{CJK})$^2$ and C. P. Sun  (\begin{CJK}{UTF8}{gbsn}孙昌璞\end{CJK})$^{1,*}$}

\affil{$^1$Graduate School of China Academy of Engineering Physics, Beijing 100193, China}

\affil{$^2$Beijing Computational Science Research Center, Beijing 100193, China}

\affil{$^*$Corresponding author}

\email{suncp@gscaep.ac.cn}

\keywords{Third quantization, Macroscopic quantum effects, BCS-BEC crossover}

\begin{abstract}
We revisit the quantization of the order parameter, which we refer to as \textsl{third quantization}, from the perspective of the commutation relation between the phase operator of the order parameter and the particle-number operator. We show that this macroscopic commutation relation does not constitute an independent fundamental postulate added to quantum mechanics, but instead emerges naturally from second quantization in the thermodynamic limit for both bosonic and fermionic many-body systems. In this sense, both Bose–Einstein condensates (BECs) and Bardeen–Cooper–Schrieffer (BCS) states can be understood as macroscopic quantum states described by bosonic coherent states: in BEC, bosons condense into a single coherent mode with a well-defined phase, while in BCS systems, collective excitations of Cooper pairs can also acquire an effectively bosonic coherent description. On this basis, we propose a new macroscopic interpretation of the BCS–BEC crossover. To characterize this crossover, we model a conventional superconductor as an assembly of macroscopically separated superconducting segments. As the intra-segment coupling increases, the system evolves from a BCS-like regime toward a BEC-like regime, in which the segments collectively behave as macroscopic coherent states. Inter-segment tunneling then locks their phases, establishes global phase coherence, and gives rise to a bulk Bose–Einstein condensate. The phase diagram of the BCS–BEC crossover can thus be understood as a manifestation of a macroscopic quantum process governed by the coherent-state dynamics of the order parameter. Our results provide a unified perspective on BEC, BCS superconductivity, and the BCS–BEC crossover within the framework of third quantization.
\end{abstract}

\section{Introduction}
Macroscopic quantum phenomena play a central role in our understanding of quantum mechanics in many-body systems. Among the most important examples are Bose–Einstein condensation (BEC) and superconductivity \cite{Bose1924,Einstein1924,Einstein1925,BCS1957}, both of which arise from spontaneous symmetry breaking and the emergence of a macroscopic order parameter with a well-defined phase \cite{Penrose_1956,Anderson_1958,Yang1962,barnett1996condensate}. In such systems, the order parameter is usually introduced at the mean-field level as a classical quantity characterizing long-range coherence. However, when particle-number fluctuations become relevant, the phase of the order parameter can no longer be regarded as a purely classical parameter and must instead be treated as a quantum dynamical variable \cite{ANDERSON_1966,Leggett2001Rewiev,tinkham2004introduction}.

This observation leads to the commutation relation between the phase operator of the order parameter and the particle-number operator. We refer to the corresponding quantization of the order parameter as \textsl{third quantization}. The terminology emphasizes that the order parameter itself emerges only after spontaneous symmetry breaking in a many-body system already described within second quantization, and is then re-quantized as a macroscopic collective degree of freedom. An important conceptual question therefore arises: is this macroscopic commutation relation a new fundamental postulate that must be added to quantum mechanics, or does it follow naturally from the underlying second-quantized many-body theory?

This question is closely related to the distinction between ordinary macroscopic quantum phenomena and macroscopic quantum coherence effects. As emphasized by Leggett \cite{leggett1980Macroscopic}, conventional macroscopic quantum states such as BECs and superconductors are characterized by spontaneous symmetry breaking and off-diagonal long-range order (ODLRO) \cite{Penrose_1956,Yang1962}, whereas macroscopic quantum coherence effects involve coherent superpositions of distinct macroscopic states \cite{leggett1980Macroscopic,Leggett_Garg,Clark1985_1,Clark1985_2,Clark1988}. In the former case, the order parameter describes the collective phase coherence of a large number of microscopic constituents; in the latter, one probes the quantum dynamics of the macroscopic order parameter itself. Understanding how this second level of quantization emerges from the microscopic theory is therefore of both conceptual and practical importance.

In this paper, we show that the commutation relation associated with third quantization can indeed be derived naturally from second quantization in the thermodynamic limit, both for bosonic condensates and for fermionic paired states. This demonstrates that third quantization is not an additional fundamental principle of quantum mechanics, but rather an emergent macroscopic structure associated with spontaneous symmetry breaking in many-body systems. From this perspective, both BEC and BCS states can be understood within a common language of bosonic coherent states. In a BEC, a macroscopic number of bosons occupy a single coherent mode with a common phase. In a BCS superconductor, the collective excitations of Cooper pairs can likewise acquire an effectively bosonic description, so that the superconducting state may also be interpreted in terms of a macroscopic coherent state of the order parameter.

This viewpoint leads naturally to a new interpretation of the BCS–BEC crossover. Conventionally, the crossover is understood as the evolution of a fermionic paired state into a condensate of tightly bound bosonic molecules as the pairing interaction increases  \cite{Leggett1980,NSR1985,Taylor2014,Chen_2024}. Here we approach the problem from the perspective of third quantization. To make the macroscopic structure explicit, we imagine that a superconductor is divided into N macroscopically separated superconducting segments. Each segment can be described as a macroscopic coherent state. As the intra-segment coupling increases, the system evolves from a BCS-like regime toward a BEC-like regime, in which the segmented superconducting states acquire increasingly bosonic collective character. Inter-segment tunneling then locks the phases of these macroscopic coherent states, establishing global phase coherence and giving rise to a bulk condensate. In this way, the BCS–BEC crossover can be viewed as a macroscopic quantum process governed by the coherent-state dynamics of the order parameter.

The main purpose of this work is therefore twofold. First, we establish third quantization as an emergent consequence of second quantization in the thermodynamic limit for both BEC and BCS systems. Second, using this framework, we provide a unified macroscopic quantum interpretation of the BCS–BEC crossover in terms of coherent-state formation, phase locking, and global condensation. This perspective connects the quantization of the order parameter directly to the evolution between two fundamental types of macroscopic quantum matter.

This paper is organized as follows. In Sec. \ref{secII}, we consider a BEC system and show, using a variational approach, that the order parameter possesses a global phase satisfying the Gross–Pitaevskii equation. We then quantize this phase and show that the phase operator and the particle-number operator satisfy a canonical commutation relation in the thermodynamic limit. In Sec. \ref{sec:BCS-BEC-condition}, we extend this analysis to superconductors and show that the phase operator of the superconducting order parameter and the Cooper-pair number operator likewise form a canonical conjugate pair. We further identify the conditions under which collective Cooper-pair excitations can be treated as bosons, and show how the superconducting ground state evolves from a BCS-like state toward a single-mode bosonic coherent state. In Sec. \ref{sec4}, we consider a superconductor divided into $N$ macroscopic segments, and analyze how global phase coherence emerges through the competition between Cooper-pair tunneling and Coulomb blockade, as well as how this coherence is characterized by macroscopic off-diagonal long-range order. Finally, in Sec. \ref{sec5}, we analyze the BCS–BEC crossover for the coupled N-segment system and in the uniform-superconductor limit, and relate the crossover behavior to the macroscopic quantum dynamics of the order parameter.

\section{Third quantization From Second quantization: Boson}\label{secII}

In this section, starting from the second quantization of the multi-atom
system, we revisit that the order parameter emerging from spontaneous
symmetry breaking exhibits global phase coherence. We further demonstrate
that, in the thermodynamic limit, the phase of the order parameter
can be re-quantized (third quantization), and the quantized
phase operator $\hat{\phi}$ and the total particle number operator
$\hat{N}$ satisfy the commutation relation: $[\hat{\phi},\hat{N}]=-i.$ 
And this phase degree of freedom corresponds to the Nambu-Goldstone
mode associated with spontaneous breaking of the global U(1) symmetry
in elemetary particle physics \cite{Goldstone_1962}.

A Bose-Einstein condensate (BEC) composed of many neutral atoms is
described by
\begin{equation}
\hat{H}=\int d\boldsymbol{r}\:[\hat{\psi}^{\dagger}(\boldsymbol{r})h(\boldsymbol{r})\hat{\psi}(\boldsymbol{r})+\frac{g}{2}\hat{\psi}^{\dagger}(\boldsymbol{r})\hat{\psi}^{\dagger}(\boldsymbol{r})\hat{\psi}(\boldsymbol{r})\hat{\psi}(\boldsymbol{r})],
\label{eq:BEC-H-1}
\end{equation}
where the single-particle Hamiltonian is $h(\boldsymbol{r}):=-\hbar^{2}\nabla^{2}/2m+V(\boldsymbol{r})$
with the constrained potential $V(\boldsymbol{r})$, and $g$ is two-body
interaction between neutral atoms, whose magnitude depends on the
$s$-wave scattering cross-section between atoms. In the coordinate
representation, the eigenequation of $h(\boldsymbol{r})$ is 
\begin{equation}
	[-\frac{\hbar^{2}}{2m}\nabla^{2}+V(\boldsymbol{r})]u_{n}(\boldsymbol{r})=E_{n}u_{n}(\boldsymbol{r}),
	\label{eq2}
\end{equation}
where $u_{n}(\boldsymbol{r})$ is the eigenfunction with corresponding
eigenenergy $E_{n}$. Accordingly, the field operator is expanded
as $\hat{\psi}(\boldsymbol{r})=\sum_{n}u_{n}(\boldsymbol{r})\hat{a}_{n}$,
where $\hat{a}_{n}$ is the annihilation operator associated with
the single-particle state. Then, the Hamiltonian (\ref{eq:BEC-H-1})
is rewritten as
\begin{equation}
	\hat{H}=\sum_{n}E_{n}\hat{a}_{n}^{\dagger}\hat{a}_{n}+\frac{1}{2}\sum_{nm,ts}g_{nm,ts}\hat{a}_{\mathrm{n}}^{\dagger}\hat{a}_{\mathrm{m}}^{\dagger}\hat{a}_{\mathrm{t}}\hat{a}_{\mathrm{s}},
\end{equation}
where $g_{nm,ts}$ is determined by the two-body interaction potential 
\begin{equation}
	g_{nm,ts}=g\int d\boldsymbol{r}\:u_{n}^{*}(\boldsymbol{r})u_{m}^{*}(\boldsymbol{r})u_{t}(\boldsymbol{r})u_{s}(\boldsymbol{r}).
	\label{eq4}
\end{equation}

We assume that the variational ground state of the interacting neutral-atom
system is a multimode coherent state:
\begin{equation}
	|\mathrm{BEC}\rangle\equiv\exp[\sum_{n}\alpha_{n}e^{i\phi_{n}}\hat{a}_{n}^{\dagger}-\alpha_{n}e^{-i\phi_{n}}\hat{a}_{n}]|\mathrm{vac}\rangle,\label{eq:BEC-gs-1}
\end{equation}
where $\alpha_{n}$ and $\phi_{n}$ are real parameters. Under the
constraint of particle-number conservation, the free energy of the
system in the ground state (\ref{eq:BEC-gs-1}) is obtained as
\begin{equation}
	F(\{\phi_{n}\},\{\alpha_{n}\}) =\langle\mathrm{BEC}|\hat{H}-\mu\hat{N}|\mathrm{BEC}\rangle,\label{eq:BEC-free-energy-1}
\end{equation}
where $\hat{N}=\int d\boldsymbol{r}\,\hat{\psi}^{\dagger}(\boldsymbol{r})\hat{\psi}(\boldsymbol{r})$,
and $\mu$ is the chemical potential. Minimizing the free energy with
the phase $\phi_{n}$ yields a common phase for all modes {[}see Appendix
\ref{sec:The-Emergence-of}{]}
\begin{equation}
	\phi_{n}=\phi+2m\pi\quad m\in\mathbb{Z}.\label{eq:BEC-equiv-phase-1}
\end{equation}
Thus, the ground state of BEC possesses a global phase: 
\begin{equation}
	|\mathrm{BEC}\rangle\equiv|\mathrm{BEC}(\phi)\rangle=\exp[\sum_{n}\alpha_{n}e^{i\phi}\hat{a}_{n}^{\dagger}-\alpha_{n}e^{-i\phi}\hat{a}_{n}]|\mathrm{vac}\rangle.\label{eq:BEC_states}
\end{equation}
Further minimization with respect to the amplitudes $\alpha_{n}$
yields the static Gross--Pitaevskii equation \cite{Shi_2018,Shi_2019,shi2019trapped_BEC},
\begin{equation}
	[-\frac{\hbar^{2}}{2m}\nabla^{2}+V(\boldsymbol{r})+g|\psi(\boldsymbol{r})|^{2}]\psi(\boldsymbol{r})=\mu\psi(\boldsymbol{r}),\label{eq:GP-eq-1}
\end{equation}
where the order parameter is defined as 
\begin{equation}
	\psi(\boldsymbol{r})\equiv e^{i\phi}\sum_{n}u_{n}(\boldsymbol{r})\alpha_{n}.\label{eq:BEC_order-parameter}
\end{equation}
For different values of $\phi$, the ground states $|\mathrm{BEC}(\phi)\rangle$ are
degenerate. A condensate corresponds to a selection of a particular
value of $\phi$, thereby breaking the U(1) symmetry and leading to
spontaneous symmetry breaking, and then the order parameter naturally
emerges {[}see Eq. (\ref{eq:BEC_order-parameter}){]}. 

In the above revisit of the BEC system, the order parameter possesses
a global phase $\phi$. We will show that the phase of the order parameter
can be re-quantized from the second quantization; namely, the phase
is quantized to be an opertor that satisfies a commutation relation
$[\hat{\phi},\hat{N}]=-i$. We then show that this commutation relation is not an additional fundamental postulate of quantum mechanics, but rather follows naturally from second quantization. In fact, $|\text{BEC}(\phi)\rangle\equiv|\phi\rangle$
can be expressed as a superposition of states with different particle
numbers \cite{ANDERSON_1966}:
\begin{equation}
	|\phi\rangle=\sum_{m=0}^{\infty}e^{im\phi}|\psi_{m}\rangle,\label{eq:expanded}
\end{equation}
where the $m$-bosons component is
\begin{equation}
|\psi_{m}\rangle=\frac{e^{-\frac{\Omega}{2}}}{m!}\sum_{n_{1},\dots,n_{m}}(\alpha_{n_{1}}\dots\alpha_{n_{m}})a_{n_{1}}^{\dagger}\dots a_{n_{m}}^{\dagger}|\text{vac}\rangle.
\end{equation}
Here, $\Omega=\sum_{n=1}^{M}\alpha_{n}^{2}$ and $M$ denotes the
total number of bosonic modes. The overlap of $\langle\phi^{\prime}|\phi\rangle$
reads
\begin{equation}
	\langle\phi^{\prime}|\phi\rangle=\prod_{n=1}^{M}\exp[-\alpha_{n}^{2}(1-e^{i(\phi-\phi^{\prime})})].
\end{equation}
It is clear that, $\langle\phi|\phi\rangle=1$. For $\phi^{\prime}\neq\phi$,
the absolute value of each factor is less than unity in the product;
hence, in the large $M$-limit ($M\to\infty$), analogous to the thermodynamic
limit, the overlap vanishes. Then one obtains the orthogonal states
$|\phi\rangle$, satisfying
\begin{equation}
	\langle\phi^{\prime}|\phi\rangle=\delta_{\phi,\phi^{\prime}}.
\end{equation}

Next, we consider the so-called third quantization, where
the phase is treated as an operator, i.e., $\phi\rightarrow\hat{\phi}$.
Let $|\{m_{i}\}\rangle=|m_{1},\dots,m_{M}\rangle$ denote the Fock
state with total particle number $m=\sum_{j=1}^{M}m_{j}$, such that
\begin{equation}
	\hat{N}|\{m_{i}\}\rangle=m|\{m_{i}\}\rangle.
\end{equation}
The particle-number operator in the $|\phi\rangle$-representation
reads
\begin{equation}
\begin{aligned}
	\langle\phi|\hat{N}|\{m_{i}\}\rangle & =\sum_{n}e^{-in\phi}\langle\psi_{n}|\hat{N}|\{m_{i}\}\rangle \\
	& =e^{-im\phi}m\langle\psi_{m}|\{m_{i}\}\rangle=i\frac{d}{d\phi}\left(\langle\phi|\{m_{i}\}\rangle\right).
\end{aligned}
\end{equation}
This shows that \cite{ANDERSON_1966}
\begin{equation}
	\hat{N}=i\frac{\partial}{\partial\phi}.
\end{equation}
In the $|\phi\rangle$-representation, the phase-factor operator is
defined as 
\begin{equation}
	e^{i\hat{\phi}}=\int_{0}^{2\pi}\frac{d\phi}{2\pi}e^{i\phi}|\phi\rangle\langle\phi|.
	\label{eq18}
\end{equation}
Thus, the phase operator and particle-number operator satisfies the
canonical commutation relation:
\begin{equation}
	[\hat{\phi},\hat{N}]=-i.
	\label{eq:commu}
\end{equation}

Obviously, in large-$M$ limit, the quantized phase operator defined here can be considered as the Pegg--Barnett phase operator. In fact,
to ensure Hermiticity, the phase operator defined by Pegg and Barnett in quantum optics must be restricted to an ($s+1$)-dimensional Fock
space of a single-mode boson. Only in the limit $s\rightarrow\infty$ does the Pegg--Barnett phase operator become defined over the entire
Fock space and thus well defined \cite{Pegg_1988,Yu_Shixi_1997,Lynch1995367}. Compared with the Pegg--Barnett phase operator, the order-parameter
phase operator defined here acts in the coherent-state space of bosons formed by the collective excitations. Specifically, the collective
excitation operator is defined by
\begin{equation}
	\hat{b}=\frac{1}{\sqrt{\Omega}}\sum_{n=1}^{M}\alpha_{n}\hat{a}_{n}^{\dagger},
\end{equation}
and BEC state is rewritten as a coherent state of a single-mode boson:
\begin{equation}
	|\phi\rangle=\exp[\sqrt{\Omega}(e^{i\phi}\hat{b}^{\dagger}-e^{-i\phi}\hat{b})]|\mathrm{vac}\rangle,
\end{equation}
where $\hat{b}$ satisfies the standard bosonic commutation relation:
\begin{equation}
	[\hat{b},\hat{b}^{\dagger}]=1.
\end{equation}
Consequently, in the large-$M$ limit $M\rightarrow\infty$, as $M\sim s\rightarrow\infty$,
the Hermiticity of the order-parameter phase operator is automatically
ensured. A more detailed derivation is provided in Appendix \ref{sec:pb-operator}.

\section{Third quantization From Second quantization: Fermion } \label{sec:BCS-BEC-condition}

Since a large number of Cooper pairs behave like a boson, a similar
consideration also applies to the condensation of Cooper pairs in
a superconductor. Thus, we will show that the superconducting order
parameter, which emerges from spontaneous symmetry breaking in a multi-electron
system, possesses a unified phase, and this phase can also be re-quantized.
We will further discuss the conditions under which the collective
excitations of Cooper pairs exhibit bosonic behavior. 

A superconductor can be described by the attractive Hubbard model \cite{NSR1985}:
\begin{equation}
	\hat{H}  =\sum_{\boldsymbol{n}}\sum_{\sigma} 3t\hat{c}_{\boldsymbol{n}\sigma}^{\dagger}\hat{c}_{\boldsymbol{n}\sigma}-t\sum_{\boldsymbol{\delta}}\left[\hat{c}_{\boldsymbol{n}\sigma}^{\dagger}\hat{c}_{\boldsymbol{n}+\boldsymbol{\delta},\sigma}+\mathrm{H.c.}\right]
	 -\sum_{\boldsymbol{n}}U\hat{c}_{\boldsymbol{n}\uparrow}^{\dagger}\hat{c}_{\boldsymbol{n}\downarrow}^{\dagger}\hat{c}_{\boldsymbol{n}\downarrow}\hat{c}_{\boldsymbol{n}\uparrow}.\label{eq:HU}
\end{equation}
Here, $\boldsymbol{n}=(n_{x},n_{y},n_{z})$ denotes the site position, $\boldsymbol{\delta}$ runs over the three nearest-neighbor directions
in three dimensions, $t$ is the hopping strength, and $U$ is the
onsite attractive interaction. The Hamiltonian in momentum space reads
\begin{equation}
	\hat{H}=\sum_{\boldsymbol{k},\sigma}\epsilon_{\boldsymbol{k}}\hat{c}_{\boldsymbol{k}\sigma}^{\dagger}\hat{c}_{\boldsymbol{k}\sigma}-U\sum_{\boldsymbol{k},\boldsymbol{q},\boldsymbol{p}}\hat{c}_{\boldsymbol{k},\uparrow}^{\dagger}\hat{c}_{-\boldsymbol{k}\downarrow}^{\dagger}\hat{c}_{\boldsymbol{q}-\boldsymbol{p},\downarrow}\hat{c}_{\boldsymbol{k}+\boldsymbol{p},\uparrow},\label{eq:Hk}
\end{equation}
where $\epsilon_{\boldsymbol{k}}=t[3-\cos(\boldsymbol{k}\cdot\boldsymbol{a})]$
and $\boldsymbol{a}$ is the  primitive lattice vector. At zero temperature, we take the
variational ground state
\begin{equation}
	|\mathrm{BCS}\rangle=\prod_{\boldsymbol{k}}(\cos\theta_{\boldsymbol{k}}+e^{i\phi_{\boldsymbol{k}}}\sin\theta_{\boldsymbol{k}}\,\hat{c}_{\boldsymbol{k}\uparrow}^{\dagger}\hat{c}_{-\boldsymbol{k}\downarrow}^{\dagger})|\mathrm{vac}\rangle
	\label{eq:BCS-state}
\end{equation}
of the many-electron system, where $\phi_{\boldsymbol{k}}$ and $\theta_{\boldsymbol{k}}$
are real variational parameters. Under the constraint of particle-number
conservation, the ground state is determined by minimizing the free
energy
\begin{equation}
	\delta F:=\delta[\langle\mathrm{BCS}|\hat{H}-\mu\hat{N}|\mathrm{BCS}\rangle]=0,
\end{equation}
where $\hat{N}=\sum_{\boldsymbol{k}}\hat{c}_{\boldsymbol{k}\sigma}^{\dagger}\hat{c}_{\boldsymbol{k}\sigma}$
is the total electron number operator. This minimization yields \cite{Kouzoudis_2010}
\begin{equation}
	\phi_{\boldsymbol{k}}=\phi+2m\pi\,(m\in\mathbb{Z})
	\label{eq:SC-phase-1}
\end{equation}
for each $\boldsymbol{k}$, and leads to the gap equation at zero temperature \cite{de2018superconductivity}
\begin{equation}
1=-\frac{U}{2}\sum_{\boldsymbol{k}}\frac{1}{\sqrt{\varepsilon_{\boldsymbol{k}}^{2}+U^2|\Delta|^{2}}},\label{eq:SC-pairing-2}
\end{equation}
where $\varepsilon_{\boldsymbol{k}}=\epsilon_{\boldsymbol{k}}-\mu$.
Here, the superconducting order parameter is defined as
\begin{equation}
	\Delta=e^{i\phi}|\Delta|\equiv-e^{i\phi}\frac{1}{2}\sum_{\boldsymbol{q}}\sin2\theta_{\boldsymbol{q}}.\label{eq:order-parameter}
\end{equation}
It follows from Eq. (\ref{eq:BCS-state}) and (\ref{eq:SC-phase-1})
that the phase of the Cooper pair with different momenta $\boldsymbol{k}$
share the same phase, implying that the superconductor possesses a
global phase. As in the BEC case, the phase of the superconducting
order parameter emerging from the second quantization is also re-quantized.
The corresponding phase operator and the operator of Cooper-pair number
$\hat{N}_{c}\equiv\hat{N}/2$ satisfy the canonical commutation relation,
as in Eq. (\ref{eq:commu}). Specifically, as in the BEC case (\ref{eq:expanded}),
$|\phi\rangle\equiv|\mathrm{BCS}(\phi)\rangle$ is rewritten as a
linear combination of states with different numbers of Cooper pairs
\cite{Anderson_1958,ANDERSON1967,tinkham2004introduction}. In the
limit where the number of Cooper-pair modes $M\rightarrow\infty$,
one readily verifies that $\langle\phi^{\prime}|\phi\rangle=\delta_{\phi^{\prime}\phi}$,
and that $\hat{N}_{c}=i\partial_{\phi}$ in the $|\phi\rangle$-representation.
This results in the canonical commutation relation $[\hat{\phi},\hat{N}_{c}]=-i$. We again demonstrate that, for fermionic systems, this commutation relation (third quantization) is not a fundamental postulate, but can be derived.
Further details are given in Appendix \ref{sec:Derivation-of-the-1}.

Since a large number of Cooper pairs behave like a boson from the second quantization, we then
examine the conditions under which the superconducting BCS state becomes
a bosonic coherent state. Historically, the connection between superconducting
states and bosonic coherent states was studied by Leggett \cite{Leggett1980}
and Nozières-Schmitt-Rink (NSR) \cite{NSR1985}, revealing how fermionic
systems can exhibit bosonic characteristics under strong attractive
interactions. Here, we revisit the crossover from the BCS state to the BEC state by demonstrating
that collective excitation operators composed of many Cooper pairs
satisfy bosonic commutation relations in the strong-interaction limit.

It follows from Eq. \eqref{eq:BCS-state} that the phase-coherent
BCS state is 
\begin{equation}
	|\phi\rangle  =\exp\left[\sum_{\boldsymbol{k}}\theta_{\boldsymbol{k}}[e^{i\phi}\hat{c}_{\boldsymbol{k}\uparrow}^{\dagger}\hat{c}_{-\boldsymbol{k}\downarrow}^{\dagger}-e^{-i\phi}\hat{c}_{-\boldsymbol{k}\downarrow}\hat{c}_{\boldsymbol{k}\uparrow}]\right]|\mathrm{vac}\rangle,
\end{equation}
where $\theta_{\boldsymbol{k}}=\arctan[\Delta U/\varepsilon_{\boldsymbol{k}}]$,
and $\phi$ is the phase of the superconducting order parameter. We
then define the annihilation operator of the collective excitation
composed of Cooper pairs as
\begin{equation}
	\hat{b}=\frac{1}{\sqrt{\Omega}}\sum_{\boldsymbol{k}}\theta_{\boldsymbol{k}}\hat{c}_{-\boldsymbol{k}\downarrow}\hat{c}_{\boldsymbol{k}\uparrow},\label{eq:collective_exication}
\end{equation}
where $\Omega=\sum_{\boldsymbol{k}}\theta_{\boldsymbol{k}}^{2}$ is
the normalization factor. The commutation relation for these collective
excitation operators is
\begin{equation}
	[\hat{b},\hat{b}^{\dagger}]=1-\hat{\eta},\label{eq:eta}
\end{equation}
where $\hat{\eta}$ quantifies deviation from the bosonic commutation
relation and is defined as
\begin{equation}
	\hat{\eta}=\frac{1}{\Omega}\sum_{\boldsymbol{k}}\theta_{\boldsymbol{k}}^{2}\hat{n}_{\boldsymbol{k}}\label{eq:nk}
\end{equation}
with $\hat{n}_{\boldsymbol{k}}$ denoting the particle number operator
for mode $\boldsymbol{k}$. 

\begin{figure}
\centering
\includegraphics[width=0.65\textwidth]{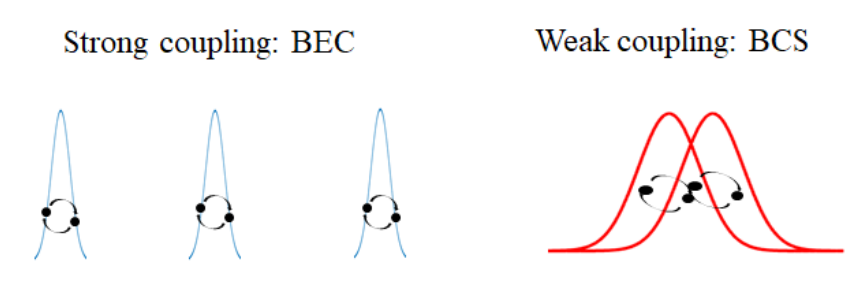}
\caption{Illustration of bound states in two limiting regimes. In the strong-interaction regime, tightly bound diatomic molecules form (left), and the system is in a bosonic coherent state. In the weak-interaction regime, Cooper pairs are spatially extended (right), forming the BCS ground state.}
\label{Fig:BEC-BCS_Picture}
\end{figure}

Actually, strong interaction ensures that the particle occupancy satisfies
$n_{\boldsymbol{k}}\ll1$, as discussed in the NSR
formalism \cite{NSR1985}. Intuitively, in the weak-interaction limit,
electrons near the Fermi surface form two-particle bound states (Cooper
pairs) \cite{Cooper1956}. These bound states are naturally spatially
extended due to the relatively weak attractive interaction (see Fig. \ref{Fig:BEC-BCS_Picture}). In this regime, the particle occupancy around and below the Fermi
surface is relatively large, and the collective behavior of Cooper
pairs deviates from bosonic characteristics {[}see Eq. (\ref{eq:eta}){]}.
In contrast, in the strong-interaction limit, the bound state becomes
tightly confined (see Fig. \ref{Fig:BEC-BCS_Picture}). Since the bound state serves as a background
for the vacuum, the particle occupancy for all modes $\boldsymbol{k}$
is low. As a result, the collective behavior of Cooper pairs behaves
like a boson.

More specifically, the expectation value of $\hat{\eta}$ in the ground
state is 
\begin{equation}
	\langle\hat{\eta}\rangle=\frac{1}{\Omega}\sum_{\boldsymbol{k}}\theta_{\boldsymbol{k}}^{2}(1-\frac{\varepsilon_{\boldsymbol{k}}}{\xi_{\boldsymbol{k}}}),
\end{equation}
where $\xi_{\mathbf{k}}=\sqrt{\varepsilon_{\mathbf{k}}^{2}+U^{2}|\Delta|^{2}}$
is the excitation energy of the quasi-particle in superconductor, and the fluctuation
of $\hat{\eta}$ is:
\begin{align}
	\Delta\hat{\eta}:=\langle\hat{\eta}^{2}\rangle-\langle\hat{\eta}\rangle^{2} & =\frac{1}{\Omega^{2}}\sum_{\boldsymbol{k}}\theta_{\boldsymbol{k}}^{4}\frac{U^{2}\Delta^{2}}{\xi_{\boldsymbol{k}}^{2}}.
\end{align}
Therefore, when the conditions $\langle\hat{\eta}\rangle\ll1$ and
$\Delta\hat{\eta}\ll1$ are satisfied, the operator $\hat{b}$ can
be approximated as a bosonic annihilation operator, satisfying the
commutation relation
\begin{equation}
	[\hat{b},\hat{b}^{\dagger}]\simeq1.
\end{equation}
Consequently, when the particle occupancy is sufficiently low (strong
interaction), the collective quasi-excitation operator (\ref{eq:collective_exication})
can be treated as a boson, and the ground state of the BCS system
becomes a coherent state with the global phase:
\begin{equation}
	|\phi\rangle=\exp\left[\sqrt{\Omega}(e^{i\phi}\hat{b}^{\dagger}-e^{-i\phi}\hat{b})\right]|\mathrm{vac}\rangle.\label{eq:Boson-coherent-state}
\end{equation}
Therefore, as the interaction strength increases, the ground state
of the superconductor gradually evolves from a BCS state to a single-mode
bosonic coherent state in Eq. (\ref{eq:Boson-coherent-state}). 
\section{Phase locking and Macroscopic off-diagnoal long range order}\label{sec4}

\begin{figure}
\centering
\includegraphics[width=0.7\textwidth]{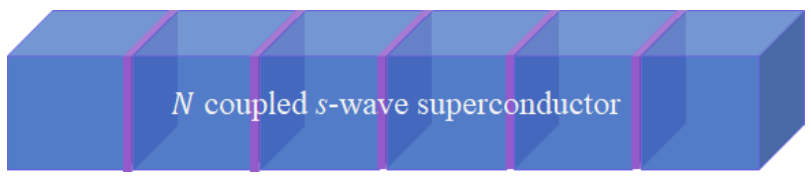} 
\caption{Schematic illustration of the $N$ coupled \textit{s}-wave superconductijng segments.}
\label{Fig2}
\end{figure}

In a superconductor, a large number of Cooper pairs behave as a single-mode
boson; accordingly, the superconducting state can be described as
a coherent state. When many superconducting segments at the macroscopic
scale are further coupled, multiple bosonic modes are realized, forming
a multimode coherent state. In this section, we investigate how the
phases of $N$ superconducting segments further lock to establish
a bulk superconducting state {[}see Fig. \ref{Fig2} (a){]}. This
system can be modeled by $\hat{H}=\sum_{j=1}^{N}\hat{H}_{j}+\hat{V}_{j,j+1},$
where the Hamiltonian of $j$-th segment is given in Eq. (\ref{eq:HU}).
In momentum space, this Hamiltonian reads as
\begin{equation}
	\hat{H}_{j}  =\sum_{\boldsymbol{k}_{j},\sigma}\varepsilon_{\boldsymbol{k}_{j}}(j)\hat{c}_{\boldsymbol{k}_{j}\sigma}^{\dagger}(j)\hat{c}_{\boldsymbol{k}_{j}\sigma}(j) -U\sum_{\boldsymbol{k}_{j},\boldsymbol{q}_{j},\boldsymbol{p}_{j}}\hat{c}_{\boldsymbol{k}_{j}\uparrow}^{\dagger}(j)\hat{c}_{\boldsymbol{q}_{j}\downarrow}^{\dagger}(j)\hat{c}_{\boldsymbol{q}_{j}-\boldsymbol{p}_{j},\downarrow}(j)c_{\boldsymbol{k}_{j}+\boldsymbol{p}_{j},\uparrow}(j).\nonumber 
\end{equation}
Here, the kinetic energy in the $j$-th segment is $\epsilon_{\boldsymbol{k}_{j}}(j)=t[3-\cos\boldsymbol{k}_{j}\cdot\boldsymbol{a}]-\mu$,
where $\boldsymbol{k}_{j}$ represents the momentum of the $j$-th
superconducting segment. The inter-segment coupling in momentum space is mediated by the Cooper-pairs tunneling \cite{Wallace_1965,Patrick_1971}
\begin{equation}
	\hat{V}_{j,j+1}=-g\sum_{\boldsymbol{k}_{j},\boldsymbol{q}_{j+1}}\hat{c}_{\boldsymbol{k}_{j}\uparrow}^{\dagger}(j)\hat{c}_{-\boldsymbol{k}_{j}\downarrow}^{\dagger}(j)\hat{c}_{-\boldsymbol{q}_{j+1}\downarrow}(j+1)\hat{c}_{\boldsymbol{q}_{j+1}\uparrow}(j+1),\label{eq:Cooper_tunneling}
\end{equation}
where $g$ is the tunneling strength for Cooper pairs.

When the Cooper-pair tunneling strength is zero $(g=0)$, each segment,
as an independent superconductor, possesses a uniform phase $\phi_{j}$.
When the tunneling strength becomes nonzero, Cooper-pair tunneling
induces fluctuations in the particle number of each segment, which
are described by $\hat{H}_{j}^{c}=-eU_{j}\hat{N}_{j}$. Here, $\hat{N}_{j}$
denotes the particle-number of the $j$-th superconducting segment,
$U_{j}$ represents the corresponding electrostatic potential, and
they satisfy the Poisson equation $\nabla^{2}U_{j}=-e\langle\hat{N}_{j}\rangle/\epsilon$, where $\epsilon$ is the dielectric permittivity.
Consequently, by taking the expectation value of $\hat{H}$ in the
product state $\prod_{j}|\phi_{j}\rangle$, the low-energy Hamiltonian
for $N$ coupled superconducting segments is obtained as [see the paper being prepared]
\begin{equation}
	H_{cla}=\sum_{j=1}^{N}E_{c}(N_{j}-N_{j+1})^{2}-E_{J}\cos(\phi_{j}-\phi_{j+1}).
\end{equation}
Here, the junction between two superconducting segments is modeled
as a parallel-plate capacitor with capacitance $C=\epsilon S/d$, where $d$ is the junction
width, and $S$ is the cross-sectional area \cite{Patrick_1971,Doniach1984}.
The corresponding charging energy is $E_{c}=e^{2}/(2C)$. The Josephson
coupling energy is $E_{J}=gU^{2}\Delta_{j}\Delta_{j+1}$, where $\Delta_{j}$
is defined in Eq. (\ref{eq:order-parameter}).

\begin{figure}
	\centering
	\includegraphics[width=0.8\textwidth]{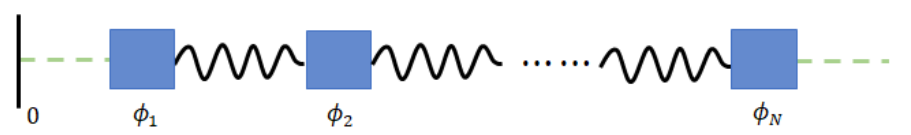}
	\caption{Schematic illustration of $N$ coupled oscillators, where the quantized phase $\phi_{j}$ of each superconducting segment serves as a coordinate-like variable.}
	\label{Fig3}
\end{figure}

Since changes in the particle number induce fluctuations in the phase
of the superconducting order parameter \cite{tinkham2004introduction},
we re-quantize the superconducting phase and the Cooper-pair number
as
\begin{equation}
	\phi_{j}\rightarrow\hat{\phi}_{j},\quad N_{c,j}\rightarrow\hat{N}_{c,j},
\end{equation}
and they satisfy the commutation relations $[\hat{\phi}_{j},\hat{N}_{c,l}]=-i\delta_{jl}$.
The resulting quantum Hamiltonian reads as \cite{Doniach1984,efetov1980phase,tinkham2004introduction,Bruder_2005}
\begin{equation}
	\hat{H}_{qua}=\sum_{j=1}^{N}4E_{c}\hat{n}_{j}-E_{J}\cos(\hat{\varphi}_{j}),
\end{equation}
where the relative phase and Cooper-pair number operators are defined
as $\hat{\varphi}_{j}=\hat{\phi}_{j}-\hat{\phi}_{j+1}$ and $\hat{n}_{j}=(\hat{N}_{c,j}-\hat{N}_{c,j+1})$.
In the $\varphi_{j}$-representation, when the phase difference satisfies
$\varphi_{j}\ll1$, the Hamiltonian expanded to second order in $\varphi_{j}$
is 
\begin{equation}
	\hat{H}_{qua}\simeq\sum_{j=1}^{N}[-16E_{c}\frac{\partial^{2}}{\partial\varphi_{j}^{2}}+\frac{E_{J}}{2}\varphi_{j}^{2}].\label{eq:oscillator_H}
\end{equation}
This Hamiltonian is equivalent to that of $N$ independent harmonic
oscillators with the effective mass $\hbar^{2}/(32E_{c})$ and frequency
$\sqrt{8E_{J}E_{c}}/\hbar$ , as shown in Fig. \ref{Fig3}. Thus,
the ground-state wave function in the $|\phi_{j}\rangle$-representation
is 
\begin{equation}
	\psi(\{\phi_{j}\})=\prod_{j=1}^{N-1}\left(\frac{E_{J}}{2\pi^{2}E_{c}}\right)^{\frac{1}{8}}\exp[-\sqrt{\frac{E_{J}}{8E_{c}}}(\phi_{j+1}-\phi_{j})^{2}]
\end{equation}
In this ground state, the average of phase difference vanishes, $\langle\hat{\varphi}_{j}\rangle=0$,
while its quantum fluctuation is $\sigma_{\varphi}=(2E_{c}/E_{J})^{\frac{1}{4}}$.  

\begin{figure}
\centering
\includegraphics[width=0.5\textwidth]{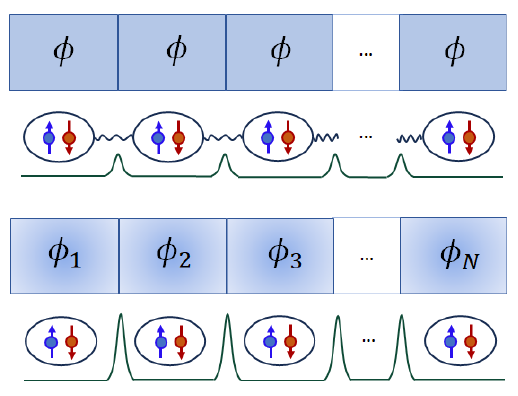}
\caption{When the Coulomb blockade of Cooper-pair tunneling dominates $(2E_{c}>E_{J})$, each superconducting segment possesses a unified phase order parameter, while the phases differ between segments (upper panel). When Cooper-pair tunneling domain $(E_{J}>2E_{c})$, the phases bewteen segments become locked, and the order parameters of all segments share a common phase, $\phi_{1}=\phi_{2}=\ldots=\phi_{N}\equiv\phi.$}
\label{Fig2-2}
\end{figure}

In the limit of $E_{J}\rightarrow0$,
the Josephson coupling vanishes and the ground-state probability density
approach $|\psi|^{2}\rightarrow$0. The superconducting segments are
thus decoupled, and each segment has an independent phase {[}see Fig.
\ref{Fig2-2}(b){]}. In contrast, in the limit of $E_{c}\rightarrow0$, the Coulomb blockade of Cooper-pair
tunneling is absent \cite{Haviland_1996,haviland2000superconducting,Haviland2001},
then the probability density of the ground state approaches $|\psi|^{2}\rightarrow\prod_{j=1}^{N-1}\delta(\phi_{j+1}-\phi_{j})$.
In this limit, the quantum fluctuations of the phase differences are
completely suppressed, Cooper-pair tunneling locks the phases, resulting
in a global phase across all superconducting segments {[}see Fig.
\ref{Fig2-2}(c){]}. That is, when  $E_{c}\rightarrow0$, the superconducting segments merge into a uniform superconductor with a global phase coherence.

To quantitatively characterize the transition from
global to local phase coherence, we examine the off-diagonal long-range
order (ODLRO) \cite{Yang1962}
\begin{align}
	\rho_{jl}(n_{j},n_{l}) & =\langle\hat{c}_{n_{j}\uparrow}^{\dagger}(j)\hat{c}_{n_{j}\downarrow}^{\dagger}(j)\hat{c}_{n_{l}\downarrow}(l)\hat{c}_{n_{l}\uparrow}(l)\rangle.
\end{align}
In the direct-product state: $\prod_{j=1}^{N}|\mathrm{BCS}(\phi_{j})\rangle$,
where each segment has an independent phase, the ODLRO factorizes
as $\rho_{jl}=\bar{\Delta}_{j}\bar{\Delta}_{l}e^{i(\phi_{j}-\phi_{l})}$
where $\bar{\Delta}_{j}=U\Delta_{j}/N_{j}$. To incorporate phase
fluctuations, we promote the phase to an operator by the third quantization. Accordingly, the ODLRO becomes an
operator, namely, the quantum ODLRO of the macroscopic units:
\begin{equation}
	\hat{\rho}_{jl}=\bar{\Delta}_{j}\bar{\Delta}_{l}e^{i(\hat{\phi}_{j}-\hat{\phi}_{l})}.
\end{equation}
In the ground state determined by Eq. \eqref{eq:oscillator_H}, macroscopic
ODLRO emerges as 
\begin{equation}
	\varrho_{jl}=\langle\psi|\hat{\rho}_{jl}|\psi\rangle=2\pi\bar{\Delta}_{j}\bar{\Delta}_{l}\exp\left[-|j-l|\sigma_{\varphi}^{2}\right].
\end{equation}
The off-diagonal long-range correlations on the macroscopic scale
decay exponentially with the distance between the superconducting
segments, consistent with the result reported in Ref. \cite{Doniach1984}.
We therefore introduce the corresponding decay factor as a characteristic
scale to identify the crossover from global to local phase coherence:\footnote{As no symmetry is broken in the transition from global to local phase
	coherence, this process is identified as a crossover.}
\begin{equation}
	\sigma_{\varphi}^{2}=\sqrt{\frac{2E_{c}}{E_{J}}}=1.  
	\label{phase_crossover}
\end{equation}
Therefore, Cooper-pair tunneling dominates over the Coulomb blockade
of Cooper-pair ($E_{J}>2E_{c}$ ), thereby further locking the phases
and establishing global phase coherence. 

\section{BCS--BEC crossover as Macroscopic Quantum Effect}\label{sec5}

We have shown that, as the intra-segment interaction strength increases,
the collective quasi-excitation of Cooper pairs in each superconducting
segment can be treated as a bosonic mode, whose ground state gradually
evolves from a BCS state into a single-mode bosonic coherent state.
For $N$ coupled superconducting segments, these bosonic modes further
couple and lock to a common phase via Cooper-pair tunneling, thereby
condensing into a multimode bosonic state, i.e., a bulk BEC. In this section, we analyze the BCS--BEC crossover for the $N$ coupled superconducting segments, as well as in the uniform-superconductor limit, from the perspective of these macroscopic quantum effects and their dependence on microscopic parameters. 

For a system of $N$ superconducting segments, each macroscopic segment
can be characterized by $\hat{H}_{j}$ in Eq. (\ref{eq:HU}). Unlike
the Cooper-pair tunneling between superconducting segments in Eq.
(\ref{eq:Cooper_tunneling}), here we consider single-particle hopping
between neighboring segments:
\begin{equation}
	\hat{H}_{j,j+1}=-\sum_{\boldsymbol{n}_{j},\boldsymbol{n}_{j+1},\sigma}G_{\boldsymbol{n}_{j},\boldsymbol{n}_{j+1,}}[\hat{c}_{\boldsymbol{n}_{j},\sigma}^{\dagger}(j)\hat{c}_{\boldsymbol{n}_{j+1},\sigma}(j+1)+\mathrm{H.c.}],
\end{equation}
where $G_{\boldsymbol{n}_{j},\boldsymbol{n}_{j+1,}}$ is hopping strength, and $\boldsymbol{n}_{j}=(n_{j,x},n_{j,y},n_{j,z})$.
In the absence of Cooper-pair blockade $E_{c}=0$, and when the hopping strength satisfies $G_{\boldsymbol{n}_{j},\boldsymbol{n}_{j+1,}}=t\delta_{n_{j,x},N_{x}}\delta_{n_{j+1,x},1}$,
the $N$ coupled segments form a bulk superconductor. In momentum
space, the single-particle tunneling Hamiltonian reads as
\begin{equation}
	\hat{H}_{j,j+1}  =-\sum_{\boldsymbol{k}_{j},\boldsymbol{q}_{j+1},\sigma}[G_{\boldsymbol{k}_{j},\boldsymbol{q}_{j+1}}c_{\boldsymbol{k}_{j}\sigma}^{\dagger}(j)c_{\boldsymbol{q}_{j+1}\sigma}(j+1)+\mathrm{H.c.}],
\end{equation}
where $G_{\boldsymbol{k}_{j},\boldsymbol{q}_{j+1}}=\sum_{\boldsymbol{n}_{j,}\boldsymbol{n}_{j+1}}G_{\boldsymbol{n}_{j},\boldsymbol{n}_{j+1,}}e^{i(\boldsymbol{k}_{j}\cdot\boldsymbol{n}_{j}-\boldsymbol{k}_{j+1}\cdot\boldsymbol{n}_{j+1})a}$
is the Fourier transform of the hopping strength. When $|G_{\boldsymbol{k}_{j},\boldsymbol{q}_{j+1}}|/(U\Delta_{j})\ll1,$
$\hat{H}_{0}=\sum_{j}\hat{H}_{j}$ is treated as the unperturbed Hamiltonian,
while the inter-segment coupling $\hat{H}_{j,j+1}$ is handled perturbatively.
By applying the Fröhlich--Nakajima transformation and retaining only
the Cooper-pair tunneling processes, Eq. (\ref{eq:Cooper_tunneling})
is obtained, where the Cooper-pair tunneling strength is \cite{Wallace_1965,Patrick_1971}
\begin{equation}
	g_{j,j+1}=\frac{G_{\boldsymbol{k}_{j},\boldsymbol{q}_{j+1}}G_{-\boldsymbol{k}_{j},-\boldsymbol{q}_{j+1}}}{\xi_{\boldsymbol{k}_{j}}+\xi_{\boldsymbol{q}_{j+1}}}\simeq\frac{G^{2}}{U[\Delta_{j}+\Delta_{j+1}]}.
\end{equation}
Here, $\xi_{\boldsymbol{k}_{j}}=\sqrt{\varepsilon_{\boldsymbol{k}_{j}}^{2}(j)+(U\Delta_{j})^{2}}$ is
the energy of quasiparticle excitation in the $j$-th superconducting
segment. From this, we will quantitatively analyze the dependence
of the BCS--BEC crossover on microscopic parameters.

\begin{figure}
\centering
\includegraphics[width=0.85\textwidth]{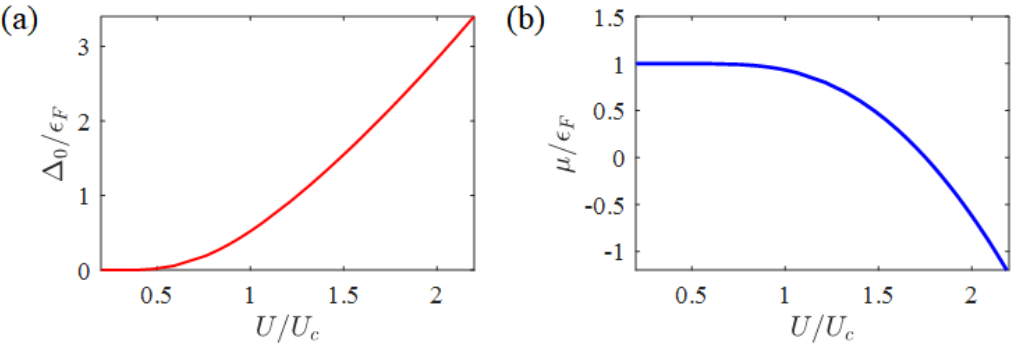}
\caption{(a) The energy gap $\Delta_0$ with varying interaction strength $U$. (b)The chemical potential $\mu$ with varying interaction strength $U$. The parameters are set as $N =k_F^3/(3\pi^2) =2\times10^{-2}k_{0}^{3}$, $U_c = (4\pi)/m k_0$, and $\epsilon_F = \hbar^2 k_F^2/(2m)$, where the cutoff momentum is $k_0 = 1.41\,\mathrm{\AA}^{-1}$. }
\label{Fig:gap_mu}
\end{figure}

At zero temperature, for a given particle number $N_{j}$ and intra-segment
interaction strength $U$ in each segment, the chemical potential and the magnitude of the order parameter are
determined self-consistently by solving the gap equation and the number equation \cite{Leggett1980,NSR1985}. To avoid the divergence problem, we use the NSR form of interaction $U_{kq}=U/\sqrt{(1+k^2/k_{0}^{2})(1+q^2/k_{0}^{2})}$, which introduce $k_0$ to constrain the interaction range. The gap equation and number equation then become
\begin{equation}
	\begin{cases}
	1=-\frac{1}{2}\sum_{\boldsymbol{k}}\frac{U}{1+\frac{k^2}{k_{0}^{2}}}\frac{1}{\sqrt{(\epsilon_{\boldsymbol{k}}-\mu)^{2}+\frac{|\Delta_{0}|^{2}}{1+k^2/k_{0}^{2}}}}, \\
		N =\sum_{\boldsymbol{k}}(1-\frac{\epsilon_{\boldsymbol{k}}-\mu}{\sqrt{(\epsilon_{\boldsymbol{k}}-\mu)^{2}+\frac{|\Delta_{0}|^{2}}{1+k^2/k_{0}^{2}}}} ).
	\end{cases}
\label{coupled_eq}
\end{equation}
Here, the subscript $j$ for the superconducting segment has been omitted. In the continuum limit, the effective electron mass is given by $m=\hbar^2/(a^2 t)$, where $a$ is the lattice constant. When $U$ exceeds a critical value $U_c= 4\pi/(m k_0)$, a bound state emerges \cite{NSR1985,Taylor2014}.  And solving Eq.~\eqref{coupled_eq} numerically gives the dependence of the chemical potential and the gap on the interaction strength. As shown in Fig. \ref{Fig:gap_mu},
as the interaction strength increases, the superconducting gap increases
monotonically, while the chemical potential gradually evolves from
positive to negative. The BCS state to bosonic coherent state for
each superconducting segment is qualitatively identified by $\mu=0$
\cite{Leggett1980}. More intuitively, the BCS--BEC crossover can
be characterized by whether the collective behavior of Cooper pairs
becomes boson-like and whether the ground state is a bosonic coherent
state. Specifically, for the $j$-th superconducting segment, the
collective excitation operator of a large number of Cooper pairs is
defined as 
\begin{equation}
	\hat{b}_{j}=\frac{1}{\sqrt{\Omega_{j}}}\sum_{\boldsymbol{k}_{j}}\theta_{\boldsymbol{k}_{j}}(j)\hat{c}_{-\boldsymbol{k}_{j},\downarrow}(j)\hat{c}_{\boldsymbol{k}_{j}\uparrow}(j),
\end{equation}
where $\Omega_{j}=\sum_{\boldsymbol{k}_{j}}\theta_{\boldsymbol{k}_{j}}^{2}$.
As shown for a single superconducting segment in Sec. \ref{sec:BCS-BEC-condition}, in the
low-occupancy limit $\langle\hat{n}_{\boldsymbol{k}_{j}}\rangle\ll1$,
one has $\langle\hat{\eta}_{j}\rangle\ll1$ and $\langle\Delta\hat{\eta}_{j}\rangle\ll1$.
Under this condition, the operator $\hat{b}_{j}$ can be approximated
as a bosonic annihilation operator satisfying $[\hat{b}_{j},\hat{b}_{l}^{\dagger}]\simeq\delta_{jl}$,
and the ground state of each segment becomes a coherent state of the
collective bosonic mode
\begin{equation}
	|\phi_{j}\rangle=\exp\left[\sqrt{\Omega_{j}}(e^{i\phi_{j}}\hat{b}_{j}^{\dagger}-e^{-i\phi_{j}}\hat{b}_{j})\right]|\mathrm{vac}\rangle.\label{eq:Boson-coherent-state-1}
\end{equation}

\begin{figure}
\centering
\includegraphics[width=0.9\textwidth]{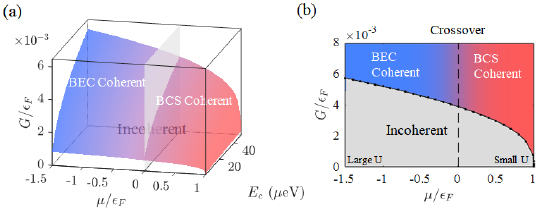}
\caption{(a) Phase diagram in the $(\mu,E_c,G)$ plane, where $\mu=0$ marks the boundary between the BCS and BEC regimes for each segment, and $E_J=2E_c$ determines the boundary between global and local phase coherence (see the color-gradient surface).  (b) shows a cross section of (a) at $E_c=50\mu \mathrm{eV}$, where the solid line denotes the boundary between the global phase coherence and local phase coherence, and the dashed vertical line at $\mu=0$ marks the boundary bewteen BCS--BEC state crossover. Parameters used here are the same as in Fig. \ref{Fig:gap_mu}.}
	\label{Fig:BCS-BEC_crossover}
\end{figure}

The $N$ superconducting segments are further coupled via Cooper-pair tunneling, and the system exhibits different phases for different tunneling strengths, as discussed in Sec.~\ref{sec4}. By Eq.~\eqref{phase_crossover} and Eq.\eqref{coupled_eq}, the phase diagram of the system is determined in the $(\mu,E_c,G)$ parameter space. As shown in Fig.~\ref{Fig:BCS-BEC_crossover}, when the tunneling strength is below the threshold so that $E_{J}<2E_{c}$, each superconducting segment has a different phase, and the system remains in a phase-incoherent state, whereas for  ($E_{J}>2E_{c}$), a global phase coherence is established among the superconducting segments. For the globally phase-coherent state, when all macroscopic segments are in coherent states of single-mode bosons, the system as a whole forms a multimode coherent state with a common phase $\phi$, thereby forming
a bulk Bose--Einstein condensate (see Fig.~\ref{Fig:BCS-BEC_crossover}). In this case, the ground state becomes
\begin{equation}
	|\phi\rangle  =\prod_{j=1}^{N}|\phi_{j}\rangle=\exp\left[\sum_{j=1}^{N}\sqrt{\Omega_{j}}(e^{i\phi}\hat{b}_{j}^{\dagger}-e^{-i\phi}\hat{b}_{j})\right]|\mathrm{vac}\rangle.
	\label{eq:multi-modes-coherence-state}
\end{equation}
By contrast, when all macroscopic segments are in the BCS state, the system as a whole is in a superconducting state (see Fig. \ref{Fig:BCS-BEC_crossover}). Remarkably, even when some segments are in the BCS state while others
are in the BEC state, the system as a whole resides in a BCS--BEC mixed state with global phase coherence. All these coherent superposition
of macroscopic state quantum phenomena deserve further investigation and attention.

When the block of Cooper-pair tunneling vanishes $E_c=0$, and the inter-segment
hopping strength satisfy $G_{\boldsymbol{n}_{j},\boldsymbol{n}_{j+1,}}=t\delta_{n_{j,x},N_{x}}\delta_{n_{j+1,x},1}$,
the inter-segment hopping strength becomes equal to the intra-segment
hopping strength. In this case, $N$ segments form a uniform superconductor.
Therefore, the BCS--BEC crossover can be intuitively understood as
a macroscopic quantum phenomenon governed by the coherent-state dynamics
of the order parameter: local phase coherence first emerges in each
bosonic mode and subsequently develops into global phase locking,
giving rise to a Bose--Einstein condensate.

\section{Conclusion}

We have investigated the quantization of the order parameter, which we refer to as \textsl{third quantization}, from the perspective of the second-quantized many-body theory. We have shown that, for both Bose–Einstein condensates and BCS superconductors, the commutation relation between the order-parameter phase and the particle-number operator can be defined in a controlled way in the thermodynamic limit. This demonstrates that third quantization is not an additional fundamental postulate of quantum mechanics, but rather an emergent macroscopic structure that follows naturally from second quantization in systems with spontaneous symmetry breaking.

With this perspective, both BEC and BCS states can be understood as macroscopic quantum states described by bosonic coherent states. In a BEC, a macroscopic number of bosons condense into a single coherent mode with a common phase. In a BCS superconductor, collective excitations of Cooper pairs can likewise acquire an effectively bosonic coherent-state description. This provides a unified perspective on two canonical forms of macroscopic quantum matter and clarifies their common foundation in the quantization of the order parameter.

Based on this viewpoint, we have further proposed a new interpretation of the BCS–BEC crossover. By modeling a conventional superconductor as an assembly of macroscopically separated superconducting segments, we showed how the system evolves from a BCS-like regime to a BEC-like regime as the intra-segment coupling increases. Each segment behaves as a macroscopic coherent state, while inter-segment tunneling locks their phases and establishes global phase coherence, eventually giving rise to a bulk condensate. In this way, the BCS–BEC crossover can be understood as a macroscopic quantum process governed by the coherent-state dynamics of the order parameter.

Our results therefore provide a unified macroscopic quantum picture of BEC, BCS superconductivity, and the BCS–BEC crossover within the framework of third quantization. More broadly, they suggest that third quantization offers a natural language for describing collective macroscopic quantum phenomena in interacting many-body systems.

In the companion paper (Paper II), we will further show that this approach can be extended from the global phase to the spatially local case, thereby providing a microscopic foundation for the quantization of superconducting transmission lines and related circuit degrees of freedom. In that sense, the present work establishes the conceptual basis, while the local-field extension developed in Paper II opens the way to applications in superconducting quantum circuits. Furthermore, within third quantization, when dissipation effects in macroscopic quantum systems are taken into account \cite{Caldeira1983,Yu_Sun_1994,Sun_Yu_1995}, it is expected that macroscopic quantum effects associated with the BCS–BEC crossover and superconducting quantum circuits may exhibit new features.

%
%


\funding{This work was supported by the Science Challenge Project (Grant No.TZ2025017),
	the National Natural Science Foundation of China (NSFC) (Grant Nos.
	12088101, 12547124), and the China Postdoctoral Science Foundation
	(Grant No. 2025M784438).}


\data{Data supporting the findings of this study are available from the authors upon reasonable request.}


\appendix

\section{The Emergence of the Order Parameter in BEC and Its Global Phase Coherence}\label{sec:The-Emergence-of}

In this section, we use a variational approach to show that the BEC order parameter possesses a uniform phase and satisfies the stationary Gross–Pitaevskii (GP) equation. 

A Bose-Einstein condensate (BEC) composed of many neutral atoms is
described by Eq. \eqref{eq:BEC-H-1}. At zero temperature, we assume that the variational ground state of the interacting neutral-atom system is a multimode coherent state:
\begin{equation}
	|\mathrm{BEC}\rangle\equiv\exp[\sum_{n}\alpha_{n}e^{i\phi_{n}}\hat{a}_{n}^{\dagger}-\alpha_{n}e^{i\phi_{n}}\hat{a}_{n}]|\mathrm{vac}\rangle
	\label{eq:BEC-gs-1-1}
\end{equation}
where $\alpha_{n}$ and $\phi_{n}$ are real parameters. Under the constraint of particle-number conservation, the free energy of the system in the ground state (\ref{eq:BEC-gs-1-1}) is obtained as
\begin{equation}
	\begin{aligned}
F&\equiv \langle\mathrm{BEC}|\hat{H}-\mu\hat{N}|\mathrm{BEC}\rangle \\
&=\sum_{n}(E_{n}-\mu)\alpha_{n}^{2}+\frac{1}{2}\sum_{nm,ts}g_{nm,ts}\alpha_{n}\alpha_{m}\alpha_{t}\alpha_{s}e^{i(\phi_{t}+\phi_{s}-\phi_{n}-\phi_{m})},
	\label{eq:Aeq2}
	\end{aligned}
\end{equation}
where $\hat{N}=\int d\boldsymbol{r},\hat{\psi}^{\dagger}(\boldsymbol{r})\hat{\psi}(\boldsymbol{r})$ is the total particle number operator, $\mu$ is the chemical potential, and $g_{nm,ts}$ is determined by the two-body interaction potential [see Eq. \eqref{eq4}]. 

Minimizing the free energy
$F(\{\phi_{n}\},\{\alpha_{n}\})$ with the phase $\phi_{n}$ yields
\begin{equation}
	\sum_{m,ts}g_{nm,ts}\alpha_{n}\alpha_{m}\alpha_{t}\alpha_{s}e^{i(\phi_{s}-\phi_{t})}\sin(\phi_{n}-\phi_{m})=0.\label{eq:BEC-phase-eq-1-1}
\end{equation}
One solution of Eq. (\ref{eq:BEC-phase-eq-1-1}) is
\begin{equation}
	\forall\,n,\quad\phi_{n}=\phi+2m\pi\quad m\in\mathbb{Z}.
	\label{eq:BEC-equiv-phase-1-1}
\end{equation}
This implies that different bosonic modes share a common phase, which yields Eq. \eqref{eq:BEC-equiv-phase-1} in the main text. Therefore, the ground state of the system possesses a global phase $\phi$, i.e., $|\mathrm{BEC}\rangle = |\mathrm{BEC}(\phi)\rangle$.
Similarly, minimizing with respect to the amplitudes $\alpha_{n}$
gives
\begin{align}
	(E_{n}-\mu)\alpha_{n}+\sum_{m,ts}g_{nm,ts}\alpha_{m}\alpha_{t}\alpha_{s}e^{i(\phi_{s}-\phi_{m})}\cos(\phi_{t}-\phi_{n}) & =0.\label{eq:BEC-pairing-eq-1-1}
\end{align}
Since the phases of different bosonic modes are coherent, Eq. (\ref{eq:BEC-pairing-eq-1-1}) simplifies to
\begin{equation}
	(E_{n}-\mu)\alpha_{\mathrm{n}}+\sum_{m,ts}g_{nm,ts}\alpha_{m}\alpha_{t}\alpha_{s}=0.\label{eq:BEC-pairing-1-1}
\end{equation}

It is easy to verify that the state $|\mathrm{BEC}(\phi)\rangle$ is the eigenstate
of the annihilation operator of the field $\hat{\psi}(\boldsymbol{r})$,
\begin{equation}
	\hat{\psi}(\boldsymbol{r})|\mathrm{BEC}(\phi)\rangle=e^{i\phi} \sum_n \alpha_n u_{n}(\boldsymbol{r})|\mathrm{BEC}(\phi)\rangle \equiv \psi(\boldsymbol{r})|\mathrm{BEC}(\phi)\rangle.
	\label{eq:BEC-order-parameter-1}
\end{equation}
The corresponding eigenfunction $\psi(\boldsymbol{r})\equiv e^{i\phi}\sum_{n}u_{n}(\boldsymbol{r})\alpha_{n}$
possesses a global phase, where $u_{n}(\boldsymbol{r})$ is the eigenfunction with corresponding
eigenenergy $E_{n}$ [see Eq. (\ref{eq2})]. This eigenfunction defines the order parameter, $\psi(\boldsymbol{r})\equiv e^{i\phi}|\psi(\boldsymbol{r})|$,
which satisfies the static Gross--Pitaevskii equation. Specifically, multiplying both sides of Eq. \eqref{eq:BEC-pairing-1-1} by $u_n(\boldsymbol{r}) e^{i\phi}$ and summing over $n$, then we have

\begin{equation}
	[-\frac{\hbar^{2}}{2m}\nabla^{2}+V(\boldsymbol{r})+g|\psi(\boldsymbol{r})|^{2}]\psi(\boldsymbol{r})=\mu\psi(\boldsymbol{r}).\label{eq:GP-eq-1-1}
\end{equation}
This corresponds to Eq.~(\ref{eq:GP-eq-1}) in the main text. It is thus demonstrated that the order parameter naturally emerges from the second quantization, and that the BEC order parameter possesses a global phase.

\section{Order-parameter and Pegg–Barnett phase operators}\label{sec:pb-operator}
In this section, we show that, in the thermodynamic limit, the phase operator of the order parameter is Hermitian, in close analogy to the Pegg–Barnett phase operator defined in quantum optics. 

In quantum optics, for a single-mode boson (e.g., a single-mode electromagnetic field), Pegg and Barnett defined a phase operator in an $(s+1)$-dimensional Hilbert space \cite{Pegg_1988}
\begin{equation}
	e^{i\hat{\theta}}=\sum_{m=0}^{s} e^{i\theta_{m}} \lvert \theta_{m} \rangle \langle \theta_{m} \rvert .
\end{equation}
where $\theta_{m}=\theta_{0}+\frac{2\pi m}{s+1},$ with $\theta_{0}$
being a reference phase. The corresponding phase states can be expanded
by the Fock states $\{|n\rangle\}$ as
\begin{equation}
\lvert\theta_{m}\rangle=\frac{1}{\sqrt{s+1}}\sum_{n=0}^{s}e^{in\theta_{m}}\lvert n\rangle.
\end{equation}
Although the above phase operator ensures Hermiticity, it must be restricted to an $(s+1)$-dimensional Fock space for a single-mode boson and is therefore not well defined. Only in the limit $s \to \infty$ does the discrete phase $\theta_m$ become continuous, and the Pegg–Barnett phase operator becomes defined over the entire Fock space and thus well defined \cite{Pegg_1988,Yu_Shixi_1997,Lynch1995367}.  In contrast to the Pegg–Barnett phase operator, the order-parameter phase operator introduced here acts in the coherent-state space of bosons formed by collective excitations. Specifically, the collective excitation operator is defined by
\begin{equation}
\hat{b}=\frac{1}{\sqrt{\Omega}}\sum_{n=1}^{M}\alpha_{n}a_{n}^{\dagger},\quad\Omega=\sum_{n=1}^{M}\alpha_{n}^{2}.
\end{equation}
From this, the BEC state can be rewritten as a coherent state of a single-mode boson 
\begin{equation}
|\phi\rangle\equiv\exp[\sqrt{\Omega}(e^{i\phi}\hat{b}^{\dagger}-e^{-i\phi}\hat{b})]|\mathrm{vac}\rangle.
\label{Aeq65}
\end{equation}
The phase operator of the order parameter defined [see Eq. \eqref{eq18}] can then be obtained through a limiting procedure as
\begin{equation}
e^{i\hat{\phi}}=\lim_{s\rightarrow\infty}\sum_{m=0}^{s}e^{i\phi_{m}}|\phi_{m}\rangle\langle\phi_{m}|,
\end{equation}
where the phase eigenstates are given by
\begin{equation}
	|\phi_{m}\rangle=\sum_{n=1}^{s}e^{i\phi_{m}n}e^{-\frac{\Omega}{2}}\frac{\Omega^{\frac{n}{2}}}{\sqrt{n!}}|n\rangle.
\end{equation}

In the BEC state given in Eq.~\eqref{Aeq65}, the average occupation number of the collective excitation is
\begin{equation}
\langle n\rangle=\sum_{n=1}^{M}\alpha_{n}^{2} =\Omega
\end{equation}
Consequently, in the thermodynamic limit $M\rightarrow\infty$, as
$M\sim s\rightarrow\infty$, the Hermiticity of the order-parameter
phase operator is automatically ensured, in accordance with the Pegg--Barnett construction \cite{Pegg_1988,Yu_Shixi_1997,Lynch1995367}.

\section{Commutation Relation of Order-Parameter Phase and Particle-Number Operators}\label{sec:Derivation-of-the-1}
In Sec. \ref{secII} of the main text, we have shown that, for a BEC system, the phase operator and particle-number operator satisfy the canonical commutation relation [see Eq.~\eqref{eq:commu}]. In this section, we show that, for a superconducting system, the commutation relation between the phase of the superconducting order parameter and the Cooper-pair number operator also satisfies $[\hat{\phi}, \hat{N}_c] = -i$, where .

It follows from Eqs. \eqref{eq:BCS-state} and \eqref{eq:SC-phase-1} that the ground state of the superconductor is 
\begin{equation}
|\text{BCS}(\phi)\rangle=\prod_{n=1}^{M}\left(\cos\theta_{k_{n}}+e^{i\phi}\sin\theta_{k_{n}}\hat{S}_{-}^{(k_{n})}\right)|\mathrm{vac}\rangle
\label{A70}
\end{equation}
where $\hat{S}_{-}^{(k_{n})}=c_{k_{n}\uparrow}^{\dagger}c_{-k_{n}\downarrow}^{\dagger}$
is the creation operator of Copper pair, and $|\mathrm{vac}\rangle$ is the
vacuum state. This ground state can be expanded as a linear combination
of states with different numbers of Cooper pairs:
\begin{equation}
|\text{BCS}(\phi)\rangle=\sum_{m=0}e^{im\phi}|\psi_{2m}\rangle,
\end{equation}
where the state with $m$ Cooper pairs is
\begin{equation}
|\psi_{2m}\rangle\equiv(\prod_{n=1}^{M}\cos\theta_{k_{n}})\frac{1}{m!}\sum_{k_{i_{1}},k_{i_{2}},\dots,k_{i_{m}}}\left(\prod_{n=1}^{M}\tan\theta_{k_{n}}\right)\hat{S}_{-}^{(k_{i_{1}})}\dots\hat{S}_{-}^{(k_{i_{m}})}|\mathrm{vac}\rangle
\end{equation}

It follows from Eq. (\ref{A70}) that the overlap $\langle\phi^{\prime}|\phi\rangle$ is
\begin{equation}
\langle\phi^{\prime}|\phi\rangle=\prod_{n=1}^{M}\left(\cos^{2}\theta_{k_{n}}+e^{i(\phi-\phi^\prime)}\sin^{2}\theta_{k_{n}}\right).
\end{equation}
Thus, when $\phi^{\prime}=\phi$, we have $\langle\phi^{\prime}|\phi\rangle=1.$
And when $\phi^{\prime}\neq\phi$, the magnitude of the overlap is:
\begin{equation}
|\langle\phi^{\prime}|\phi\rangle|=\prod_{n=1}^{M}\left|\cos^{2}\theta_{k_{n}}+e^{i(\phi-\phi')}\sin^{2}\theta_{k_{n}}\right|.
\label{eqA73}
\end{equation}
Since each factor in Eq.~(\ref{eqA73}) is less than or equal to 1 for any $\theta_{k_n}$, it follows that in the large-$M$ limit $M \to \infty$, the overlap vanishes.
\begin{equation}
|\langle\phi^{\prime}|\phi\rangle|\to0.
\end{equation}
We therefore obtain
\begin{equation}
\langle\phi^{\prime}|\phi\rangle=\delta_{\phi^{\prime}\phi}.
\end{equation}

In the $|\phi\rangle$-representation, the particle-number operator is

\begin{equation}
\langle \phi |\hat{N}|n \rangle=\sum_{m=0} e^{-im\phi} \langle \psi_{2m} |\hat{N}|n \rangle=\sum_{m=0} e^{-im\phi} n \langle \psi_{2m}|n\rangle,
\label{eq:overlap}
\end{equation}
where $|n\rangle=|n_{k_{1},}n_{k_{2}},\ldots,n_{k_{M}}\rangle$ is the $n$-particle Fock state. It is easy to verify that $\langle \psi_{2m} |n\rangle=0$ for odd $n$, while for even $n$, by taking $m=n/2$, we obtain:
\begin{equation}
\langle n|\hat{N}|\phi\rangle=2i\frac{\partial}{\partial \phi }(e^{-in\phi/2} \langle n|\psi_{n}\rangle)=2i\frac{\partial}{\partial \phi }\langle n|\phi\rangle,
\end{equation}
Thus, in the $|\phi\rangle$ representation, the particle-number operator  is
\begin{equation}
\hat{N}=2i\frac{\partial}{\partial\phi}
\end{equation}
From this result, the commutation relation between the phase operator and the Cooper-pair number operator satisfy
\begin{equation}
[\hat{\phi},\hat{N}_c]=-i
\end{equation}
where $\hat{N}_c=\hat{N}/2$ is the Cooper-pair number operator.

For a Josephson junction consisting of two superconductors, the phase and Cooper-pair number operators of each superconductor satisfy
\begin{equation}
[\hat{\phi}_{1},\hat{N}_{c,1}]=-i,\quad[\hat{\phi}_{2},\hat{N}_{c,2}]=-i
\end{equation}
Defining the phase difference $\hat{\phi}=\hat{\phi}_{1}-\hat{\phi}_{2}$
and the Cooper-pair number difference $\hat{N}_{c}=(\hat{N}_{c,1}-\hat{N}_{c,2})/2$, the corresponding commutation relation is obtained as \begin{equation}
[\hat{\phi},\hat{N}_c]=-i
\end{equation}
This result provides the theoretical foundation for circuit quantization in superconducting quantum computing.

\bibliographystyle{vancouver}
\bibliography{RefpgTQv2}

\begin{thebibliography}{10}

\bibitem{Bose1924}
Bose.
\newblock Plancks Gesetz und Lichtquantenhypothese.
\newblock Zeitschrift für Physik. 1924;26:178--181.
\newblock Available from: \url{https://doi.org/10.1007/BF01327326}.

\bibitem{Einstein1924}
Einstein A.
\newblock Quantentheorie des einatomigen idealen Gases I.
\newblock Sitzungsberichte der Preussischen Akademie der Wissenschaften. 1924;.

\bibitem{Einstein1925}
Einstein A.
\newblock Quantentheorie des einatomigen idealen Gases II.
\newblock Sitzungsberichte der Preussischen Akademie der Wissenschaften. 1925;.

\bibitem{BCS1957}
Bardeen J, Cooper LN, Schrieffer JR.
\newblock Theory of Superconductivity.
\newblock Phys Rev. 1957 Dec;108:1175--1204.
\newblock Available from:
  \url{https://link.aps.org/doi/10.1103/PhysRev.108.1175}.

\bibitem{Penrose_1956}
Penrose O, Onsager L.
\newblock Bose-Einstein Condensation and Liquid Helium.
\newblock Phys Rev. 1956 Nov;104:576--584.
\newblock Available from:
  \url{https://link.aps.org/doi/10.1103/PhysRev.104.576}.

\bibitem{Anderson_1958}
Anderson PW.
\newblock Random-Phase Approximation in the Theory of Superconductivity.
\newblock Phys Rev. 1958 Dec;112:1900--1916.
\newblock Available from:
  \url{https://link.aps.org/doi/10.1103/PhysRev.112.1900}.

\bibitem{Yang1962}
Yang CN.
\newblock Concept of Off-Diagonal Long-Range Order and the Quantum Phases of
  Liquid He and of Superconductors.
\newblock Rev Mod Phys. 1962 Oct;34:694--704.
\newblock Available from:
  \url{https://link.aps.org/doi/10.1103/RevModPhys.34.694}.

\bibitem{barnett1996condensate}
Barnett S, Burnett K, Vaccaro J.
\newblock Why a condensate can be thought of as having a definite phase.
\newblock Journal of research of the National Institute of Standards and
  Technology. 1996;101(4):593.
\newblock Available from:
  \url{https://pmc.ncbi.nlm.nih.gov/articles/PMC4907620/}.

\bibitem{ANDERSON_1966}
ANDERSON PW.
\newblock Considerations on the Flow of Superfluid Helium.
\newblock Rev Mod Phys. 1966 Apr;38:298--310.
\newblock Available from:
  \url{https://link.aps.org/doi/10.1103/RevModPhys.38.298}.

\bibitem{Leggett2001Rewiev}
Leggett AJ.
\newblock Bose-Einstein condensation in the alkali gases: Some fundamental
  concepts.
\newblock Rev Mod Phys. 2001 Apr;73:307--356.
\newblock Available from:
  \url{https://link.aps.org/doi/10.1103/RevModPhys.73.307}.

\bibitem{tinkham2004introduction}
Tinkham M.
\newblock Introduction to superconductivity.
\newblock Courier Corporation; 2004.

\bibitem{leggett1980Macroscopic}
Leggett AJ.
\newblock Macroscopic quantum systems and the quantum theory of measurement.
\newblock Progress of Theoretical Physics Supplement. 1980;69:80--100.
\newblock Available from: \url{https://doi.org/10.1143/PTP.69.80}.

\bibitem{Leggett_Garg}
Leggett AJ, Garg A.
\newblock Quantum mechanics versus macroscopic realism: Is the flux there when
  nobody looks?
\newblock Phys Rev Lett. 1985 Mar;54:857--860.
\newblock Available from:
  \url{https://link.aps.org/doi/10.1103/PhysRevLett.54.857}.

\bibitem{Clark1985_1}
Martinis JM, Devoret MH, Clarke J.
\newblock Energy-Level Quantization in the Zero-Voltage State of a
  Current-Biased Josephson Junction.
\newblock Phys Rev Lett. 1985 Oct;55:1543--1546.
\newblock Available from:
  \url{https://link.aps.org/doi/10.1103/PhysRevLett.55.1543}.

\bibitem{Clark1985_2}
Devoret MH, Martinis JM, Clarke J.
\newblock Measurements of Macroscopic Quantum Tunneling out of the Zero-Voltage
  State of a Current-Biased Josephson Junction.
\newblock Phys Rev Lett. 1985 Oct;55:1908--1911.
\newblock Available from:
  \url{https://link.aps.org/doi/10.1103/PhysRevLett.55.1908}.

\bibitem{Clark1988}
Clarke J, Cleland AN, Devoret MH, Esteve D, Martinis JM.
\newblock Quantum Mechanics of a Macroscopic Variable: The Phase Difference of
  a Josephson Junction.
\newblock Science. 1988;239(4843):992--997.
\newblock Available from:
  \url{https://www.science.org/doi/abs/10.1126/science.239.4843.992}.

\bibitem{Leggett1980}
Leggett AJ.
\newblock Diatomic Molecules and Cooper Pairs.
\newblock Journal de Physique Colloques. 1980;41(C7):C7--19--C7--26.

\bibitem{NSR1985}
Nozi{\`e}res P, Schmitt-Rink S.
\newblock Bose condensation in an attractive fermion gas: From weak to strong
  coupling superconductivity.
\newblock Journal of Low Temperature Physics. 1985;59(3-4):195--211.

\bibitem{Taylor2014}
Randeria M, Taylor E.
\newblock Crossover from Bardeen-Cooper-Schrieffer to Bose-Einstein
  Condensation and the Unitary Fermi Gas [Journal Article].
\newblock Annual Review of Condensed Matter Physics. 2014;5(Volume 5,
  2014):209--232.
\newblock Available from:
  \url{https://www.annualreviews.org/content/journals/10.1146/annurev-conmatphys-031113-133829}.

\bibitem{Chen_2024}
Chen Q, Wang Z, Boyack R, Yang S, Levin K.
\newblock When superconductivity crosses over: From BCS to BEC.
\newblock Rev Mod Phys. 2024 May;96:025002.
\newblock Available from:
  \url{https://link.aps.org/doi/10.1103/RevModPhys.96.025002}.

\bibitem{Goldstone_1962}
Goldstone J, Salam A, Weinberg S.
\newblock Broken Symmetries.
\newblock Phys Rev. 1962 Aug;127:965--970.
\newblock Available from:
  \url{https://link.aps.org/doi/10.1103/PhysRev.127.965}.

\bibitem{Shi_2018}
Shi T, Demler E, {Ignacio Cirac} J.
\newblock Variational study of fermionic and bosonic systems with non-Gaussian
  states: Theory and applications.
\newblock Annals of Physics. 2018;390:245--302.
\newblock Available from:
  \url{https://www.sciencedirect.com/science/article/pii/S0003491617303251}.

\bibitem{Shi_2019}
Guaita T, Hackl L, Shi T, Hubig C, Demler E, Cirac JI.
\newblock Gaussian time-dependent variational principle for the Bose-Hubbard
  model.
\newblock Phys Rev B. 2019 Sep;100:094529.
\newblock Available from:
  \url{https://link.aps.org/doi/10.1103/PhysRevB.100.094529}.

\bibitem{shi2019trapped_BEC}
Shi T, Pan J, Yi S. Trapped Bose-Einstein Condensates with Attractive s-wave
  Interaction; 2019.
\newblock Available from: \url{https://arxiv.org/abs/1909.02432}.

\bibitem{Pegg_1988}
Pegg DT, Barnett SM.
\newblock Unitary Phase Operator in Quantum Mechanics.
\newblock Europhysics Letters. 1988 jul;6(6):483.
\newblock Available from: \url{https://doi.org/10.1209/0295-5075/6/6/002}.

\bibitem{Yu_Shixi_1997}
Yu S.
\newblock Quantized Phase Difference.
\newblock Phys Rev Lett. 1997 Aug;79:780--783.
\newblock Available from:
  \url{https://link.aps.org/doi/10.1103/PhysRevLett.79.780}.

\bibitem{Lynch1995367}
Lynch R.
\newblock The quantum phase problem: a critical review.
\newblock Physics Reports. 1995;256(6):367--436.
\newblock Available from:
  \url{https://www.sciencedirect.com/science/article/pii/037015739400095K}.

\bibitem{Kouzoudis_2010}
Kouzoudis D.
\newblock Proof of the phase coherence in the Bardeen–Cooper–Schrieffer
  theory of superconductivity from first principles.
\newblock European Journal of Physics. 2009 dec;31(1):239.
\newblock Available from: \url{https://dx.doi.org/10.1088/0143-0807/31/1/021}.

\bibitem{de2018superconductivity}
De~Gennes PG.
\newblock Superconductivity of metals and alloys.
\newblock CRC press; 2018.

\bibitem{ANDERSON1967}
Anderson PW.
\newblock Chapter I The Josephson Effect and Quantum Coherence Measurements in
  Superconductors and Superfluids.
\newblock vol.~5 of Progress in Low Temperature Physics. Elsevier; 1967. p.
  1--43.
\newblock Available from:
  \url{https://www.sciencedirect.com/science/article/pii/S0079641708601195}.

\bibitem{Cooper1956}
Cooper LN.
\newblock Bound Electron Pairs in a Degenerate Fermi Gas.
\newblock Physical Review. 1956;104(4):1189--1190.

\bibitem{Wallace_1965}
Wallace PR, Stavn MJ.
\newblock QUASI-SPIN TREATMENT OF JOSEPHSON TUNNELING BETWEEN SUPERCONDUCTORS.
\newblock Canadian Journal of Physics. 1965;43(3):411--417.
\newblock Available from: \url{https://doi.org/10.1139/p65-037}.

\bibitem{Patrick_1971}
Lee PA, Scully MO.
\newblock Theory of Josephson Radiation. I. General Theory.
\newblock Phys Rev B. 1971 Feb;3:769--779.
\newblock Available from:
  \url{https://link.aps.org/doi/10.1103/PhysRevB.3.769}.

\bibitem{Doniach1984}
Bradley RM, Doniach S.
\newblock Quantum fluctuations in chains of Josephson junctions.
\newblock Phys Rev B. 1984 Aug;30:1138--1147.
\newblock Available from:
  \url{https://link.aps.org/doi/10.1103/PhysRevB.30.1138}.

\bibitem{efetov1980phase}
Efetov KB.
\newblock Phase transition in granulated superconductors.
\newblock Sov Phys-JETP. 1980;51(5).

\bibitem{Bruder_2005}
Bruder C, Fazio R, Schön G.
\newblock The Bose-Hubbard model: from Josephson junction arrays to optical
  lattices.
\newblock Annalen der Physik. 2005;517(9-10):566--577.
\newblock Available from:
  \url{https://onlinelibrary.wiley.com/doi/abs/10.1002/andp.200551709-1005}.

\bibitem{Haviland_1996}
Haviland DB, Delsing P.
\newblock Cooper-pair charge solitons: The electrodynamics of localized charge
  in a superconductor.
\newblock Phys Rev B. 1996 Sep;54:R6857--R6860.
\newblock Available from:
  \url{https://link.aps.org/doi/10.1103/PhysRevB.54.R6857}.

\bibitem{haviland2000superconducting}
Haviland DB, Andersson K, {\AA}gren P.
\newblock Superconducting and insulating behavior in one-dimensional Josephson
  junction arrays.
\newblock Journal of Low Temperature Physics. 2000;118(5):733--749.

\bibitem{Haviland2001}
Haviland DB, Andersson K, {\AA}gren P, Johansson J, Schöllmann V, Watanabe M.
\newblock Quantum phase transition in one-dimensional Josephson junction
  arrays.
\newblock Physica C: Superconductivity. 2001;352(1):55--60.
\newblock Available from:
  \url{https://www.sciencedirect.com/science/article/pii/S0921453400016750}.

\bibitem{Caldeira1983}
Caldeira AO, Leggett AJ.
\newblock Quantum tunnelling in a dissipative system.
\newblock Annals of Physics. 1983;149(2):374--456.
\newblock Available from:
  \url{https://www.sciencedirect.com/science/article/pii/0003491683902026}.

\bibitem{Yu_Sun_1994}
Yu LH, Sun CP.
\newblock Evolution of the wave function in a dissipative system.
\newblock Phys Rev A. 1994 Jan;49:592--595.
\newblock Available from:
  \url{https://link.aps.org/doi/10.1103/PhysRevA.49.592}.

\bibitem{Sun_Yu_1995}
Sun CP, Yu LH.
\newblock Exact dynamics of a quantum dissipative system in a constant external
  field.
\newblock Phys Rev A. 1995 Mar;51:1845--1853.
\newblock Available from:
  \url{https://link.aps.org/doi/10.1103/PhysRevA.51.1845}.

\end{thebibliography}

\end{document}